\documentclass[a4paper,11pt]{article}
\usepackage{pos}
\makeatletter
\gdef\PoScopyright@box{\parbox[b]{.7\textwidth}{\tiny $\copyright$ Copyright 2024 CERN for the benefit of the ATLAS and CMS Collaborations.\\ Reproduction of this article or parts of it is allowed as specified in the CC-BY-4.0 license.}}
\makeatother
\def\logo{}
\def\posurl{}

\title{Prompt searches for feebly interacting particles\\ at the LHC}
\ShortTitle{Prompt searches for feebly interacting particles at the LHC}

\author*{Joscha Knolle}
\onbehalf{on behalf of the ATLAS and CMS Collaborations}
\affiliation{Universiteit Gent, Ghent, Belgium}
\emailAdd{joscha.knolle@cern.ch}

\abstract{%
    Recent results from the ATLAS and CMS experiments in searches for prompt signatures of feebly interacting particles are presented.
    All presented results are based on the 2015--2018 data set of 13\,\TeV{} proton-proton collisions, corresponding to an integrated luminosity of about 140\,fb\textsuperscript{--1}.
    The discussed models include dark mesons, heavy neutral leptons, dark matter, and dark photons.
    The obtained exclusion limits significantly extend the probed parameter space and, in some cases, provide the first collider-based constraints for the considered models.
}

\FullConference{%
    12th Large Hadron Collider Physics Conference (LHCP2024)\\
    3--7 June 2024 \\
    Boston, USA
}

\graphicspath{{../../images/}{../../templates/}}

\newcommand{\TeV}{\mbox{Te\hspace{-.08em}V}}
\newcommand{\GeV}{\mbox{Ge\hspace{-.08em}V}}

\begin{document}

\maketitle

\section{Introduction}

The ATLAS~\cite{ATLAS:Detector-2008, ATLAS:Detector-2024} and CMS~\cite{CMS:Detector-2008, CMS:Detector-2024} experiments at the CERN Large Hadron Collider (LHC) have each collected about 140\,fb\textsuperscript{--1} of proton-proton collision data at a centre-of-mass energy of 13\,\TeV{} in 2015--2018.
With the large data set and high energy, a wide range of possible new physics beyond the standard model (SM) is accessible.
The search programmes of the two collaborations reflect this and cover various possible signatures motivated by theoretical models that try to address the several open questions existing in particle physics.

One class of searches considers prompt signatures of feebly interacting particles.
``Prompt'' means that the particles are produced and decay at the interaction point, in contrast to dedicated searches for long-lived particles resulting in displaced signatures, as reviewed in Ref.~\cite{LHCP:LLP}.
``Feebly interacting''~\cite{FIPs:2022} refers to the strength of the coupling with SM particles.
Strongly coupled models are often already excluded for mass values in the \GeV{} range, and searches instead focus on \TeV-scale signatures as reviewed in Refs.~\cite{LHCP:TeV1, LHCP:TeV2}.
Feebly interacting new physics can still be searched for in lower mass ranges.
While such classifications of new-physics searches are useful for an overview of the LHC search programme, an exact distinction is generally not possible and would always be ambiguous.

In this contribution, I review several recent results~\cite{ATLAS:EXOT-2023-09, CMS:EXO-22-011, CMS:EXO-20-006, ATLAS:EXOT-2018-62, CMS:EXO-21-012, ATLAS:EXOT-2018-64, CMS:EXO-21-005, ATLAS:EXOT-2023-01} by the ATLAS and CMS Collaborations that can be classified as prompt searches for feebly-interacting particles.
A more complete overview of the ATLAS and CMS search programmes is provided in a series of recent review articles by the two collaborations~\cite{ATLAS:Review-EXO, ATLAS:Review-ExoHiggs, CMS:Review-DM, CMS:Review-HNL}.

\section{Recent results}

\begin{figure}[!t]
\centering
\raisebox{3pt}{\includegraphics[width=0.455\textwidth]{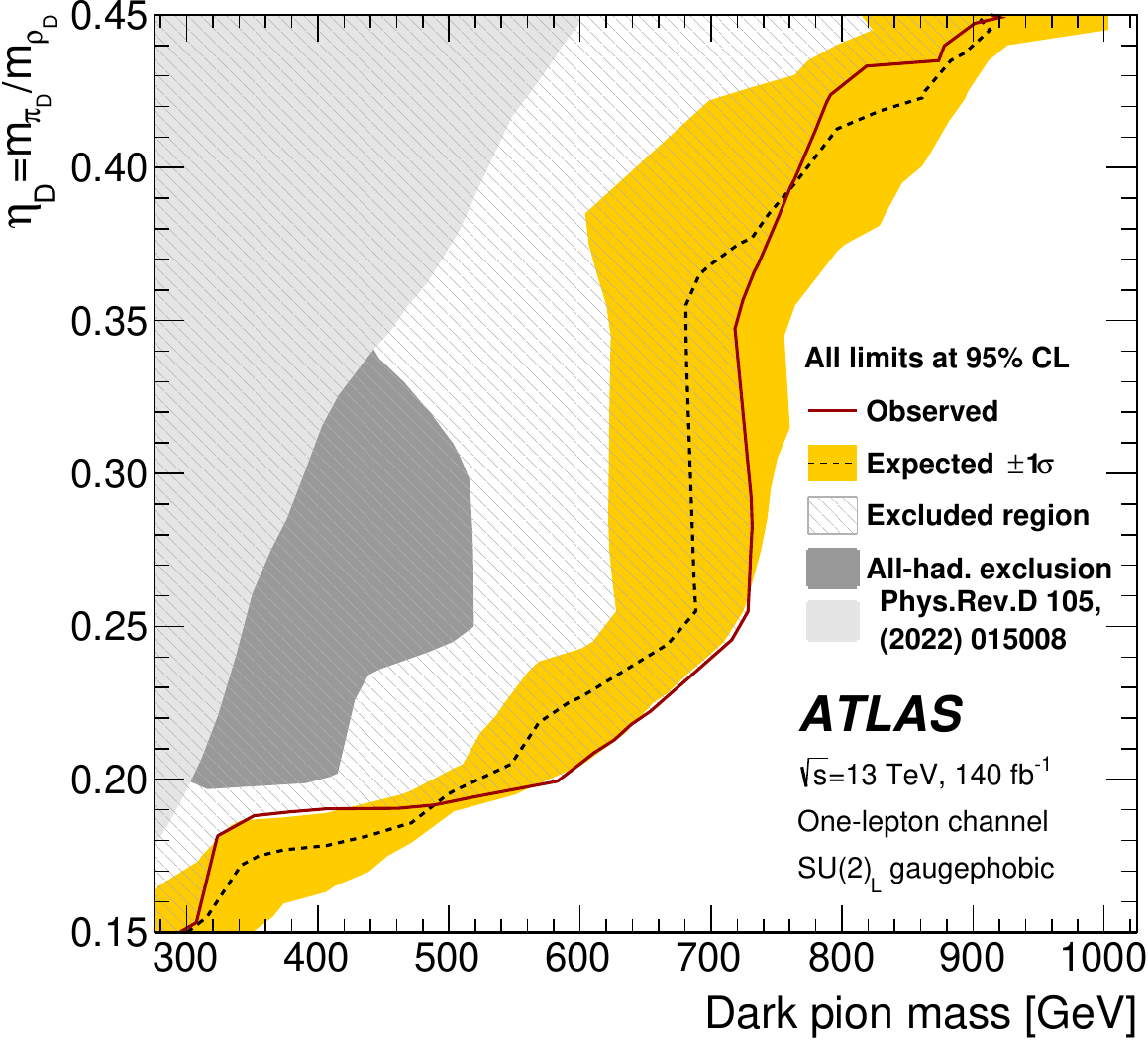}}%
\hspace*{0.05\textwidth}%
\includegraphics[width=0.445\textwidth]{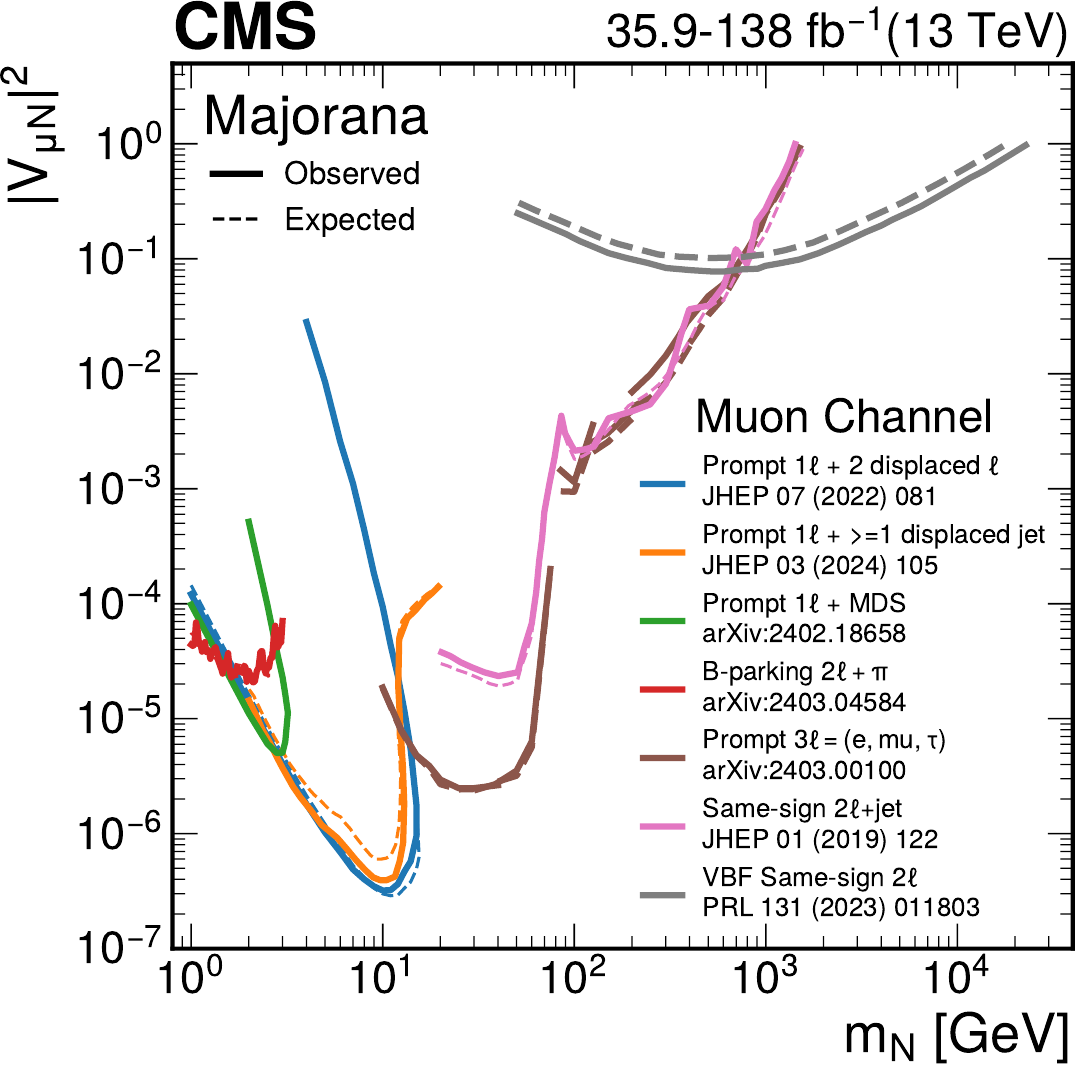}%
\caption{%
    Left: Dark meson limits on the dark-pion-to-dark-rho-meson mass ratio as a function of the dark pion mass, obtained by ATLAS in the one-lepton channel~\cite{ATLAS:EXOT-2023-09}.
    Right: HNL limits on the HNL-to-muon-neutrino mixing parameter as a function of the HNL mass, obtained by CMS in various analyses for the Majorana nature~\cite{CMS:Review-HNL}.
}
\label{fig:1}
\end{figure}

\paragraph{Dark mesons decaying to top and bottom quarks (ATLAS)~\cite{ATLAS:EXOT-2023-09}}
In stealth dark matter models, dark mesons can arise that are only weakly coupled to the SM sector but are allowed to decay into pure SM states.
This search is aimed at pair-produced dark pseudoscalar mesons (``dark pions''), where the dominant production mechanism is via a dark vector meson (``dark rho'') resonance, in final states with either three top quarks and one bottom quark, or two top quarks and two bottom quarks.
Events are selected in the all-hadronic channel with no charged leptons and at least six jets, and in the one-lepton channel with exactly one charged lepton and at least five jets.
The selected jets and lepton are reclustered to two large-radius jets, providing dark pion candidates.
Exclusion limits are derived for dark pion masses up to 940\,\GeV, considering ranges in the mass ratio between dark pion and dark rho meson of 0.15--0.45, as shown in Fig.~\ref{fig:1}~(left).
This is the first time that this type of model has been searched for at a collider experiment.

\paragraph{Heavy neutral leptons (HNLs) in prompt trilepton events (CMS)~\cite{CMS:EXO-22-011}}
Motivated by the observation of neutrino oscillations and the conclusion that neutrinos must have a nonzero mass, a simple SM extension with one Majorana or Dirac HNL mixed with one single SM neutrino generation is considered with masses between 10\,\GeV{} and 1.5\,\TeV.
This search is aimed at HNL production in association with a charged lepton, with a subsequent decay of the HNL to two charged leptons and one SM neutrino.
Events are thus selected with exactly three charged leptons, including for the first time up to one hadronically decaying $\tau$ lepton.
In separate categories for all lepton flavour combinations, dedicated strategies are applied for different target HNL mass ranges, including event categorization and multivariate analysis discriminants.
Exclusion limits are shown in Fig.~\ref{fig:1}~(right) for exclusive muon neutrino coupling, illustrating the complementarity with lower-mass HNL searches optimized for displaced signatures~\cite{CMS:EXO-20-009, CMS:EXO-21-013, CMS:EXO-21-011}, and a higher-mass HNL search in a different production channel~\cite{CMS:EXO-21-003}.
Exclusive $\tau$ neutrino couplings are probed for HNL masses above the W~boson mass for the first time.

\begin{figure}[!t]
\centering
\raisebox{2pt}{\includegraphics[width=0.34\textwidth]{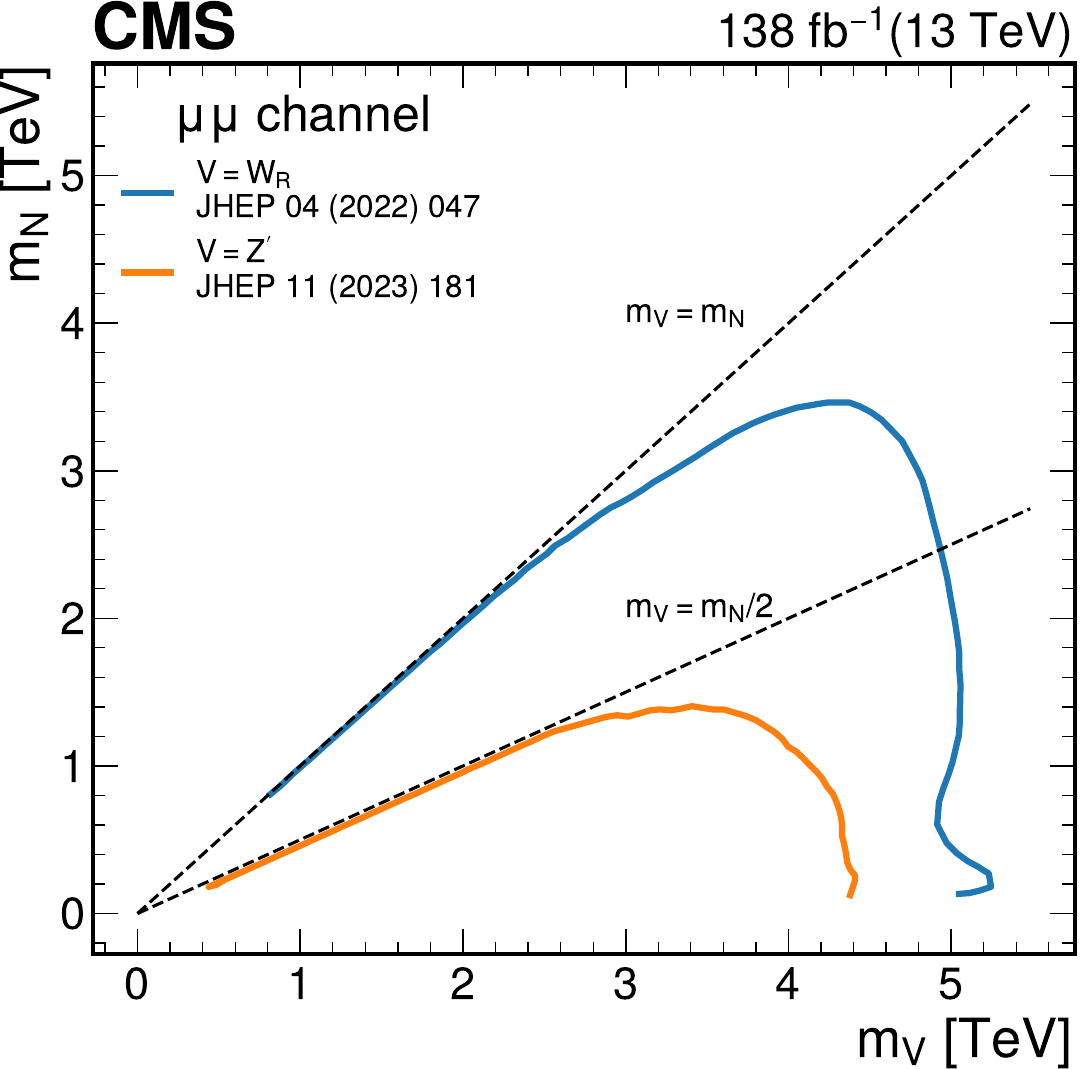}}%
\hfill%
\includegraphics[width=0.61\textwidth]{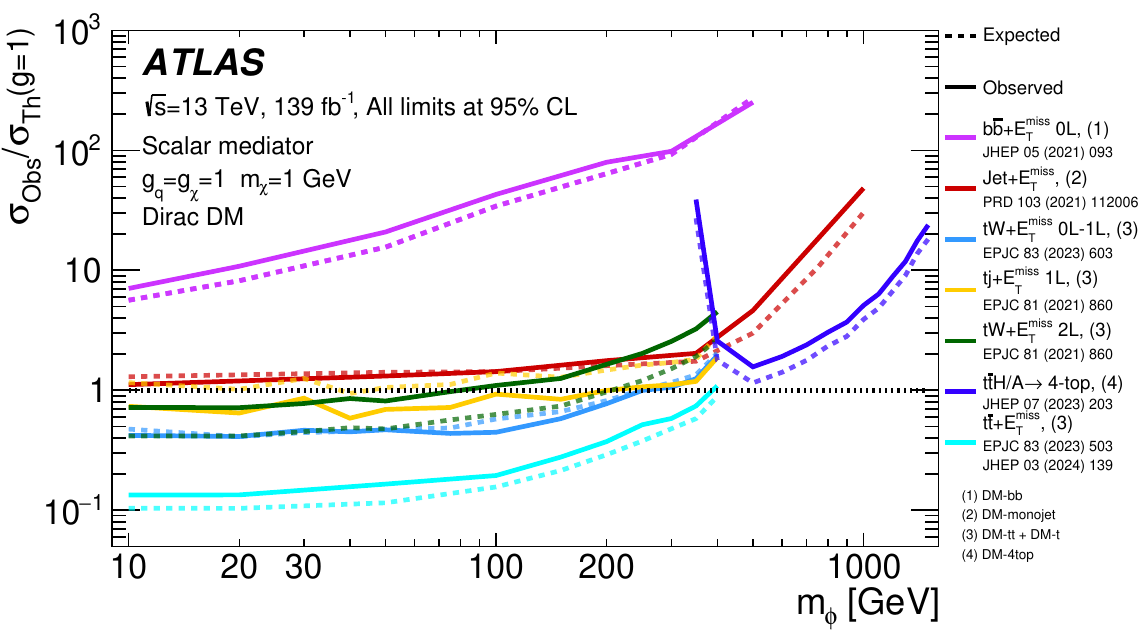}%
\caption{%
    Left: Left-right symmetry model limits on the HNL mass as a function of the new vector boson mass, obtained by CMS in two analyses in the muon channel~\cite{CMS:Review-HNL}.
    Right: Scalar DM mediator limits on the cross section relative to the predicted value as a function of the scalar mediator mass, obtained by ATLAS in various search channels~\cite{ATLAS:EXOT-2018-62}.
}
\label{fig:2}
\end{figure}

\paragraph{$\boldsymbol{\mathrm{Z}^\prime}$ bosons decaying to HNL pairs (CMS)~\cite{CMS:EXO-20-006}}
A more complete description of HNLs is provided in a left-right symmetry model that extends the SM additionally by new heavy weak gauge bosons.
This search targets production of a heavy neutral $\mathrm{Z}^\prime$ boson with decay to two Majorana HNLs, with further decays to two same-sign leptons and four quarks.
A large mass gap between the $\mathrm{Z}^\prime$ boson and the HNL is considered, resulting in boosted HNL decays.
Events are selected with two same-sign electrons or muons and additional jets, and HNL candidates are reconstructed by combining two small-radius jets with a charged lepton or by using one large-radius jet.
Fits to the reconstructed $\mathrm{Z}^\prime$ boson mass are used to derive exclusion limits for $\mathrm{Z}^\prime$ boson masses in the range 0.4--4.6\,\TeV, considering HNL masses of 100\,\GeV{} or higher.
In Fig.~\ref{fig:2}~(left), the exclusion limits are compared to the results obtained in a different production channel.

\paragraph{Simplified dark matter (DM) models with \textit{s}-channel mediator (ATLAS)~\cite{ATLAS:EXOT-2018-62}}
A common benchmark for the comparison of the sensitivity of different DM searches are simplified models with a single DM candidate particle that can be pair-produced through decays of an $s$-channel massive scalar or vector mediator.
This interpretation considers various searches targeted either at SM particles produced together with undetected DM particles resulting in large missing transverse momentum (``X\,+\,$p_{\mathrm{T}}^{\text{miss}}$'') or fully visible final states modified through the presence of the mediator.
For a scalar mediator, the strongest constraints are provided by a $\mathrm{t}\bar{\mathrm{t}}$\,+\,$p_{\mathrm{T}}^{\text{miss}}$~\cite{ATLAS:SUSY-2023-22} and a four top quark production~\cite{ATLAS:EXOT-2019-26} search, and the corresponding exclusion limits are shown in Fig.~\ref{fig:2}~(right).
For a vector mediator, the strongest constraints are provided by a $\gamma$\,+\,$p_{\mathrm{T}}^{\text{miss}}$ (``mono-photon'') search~\cite{ATLAS:EXOT-2018-63} and dijet resonance searches.
Additionally, the complementarity between the ATLAS constraints and direct DM detection experiments are evaluated.

\begin{figure}[!t]
\centering
\includegraphics[width=0.45\textwidth]{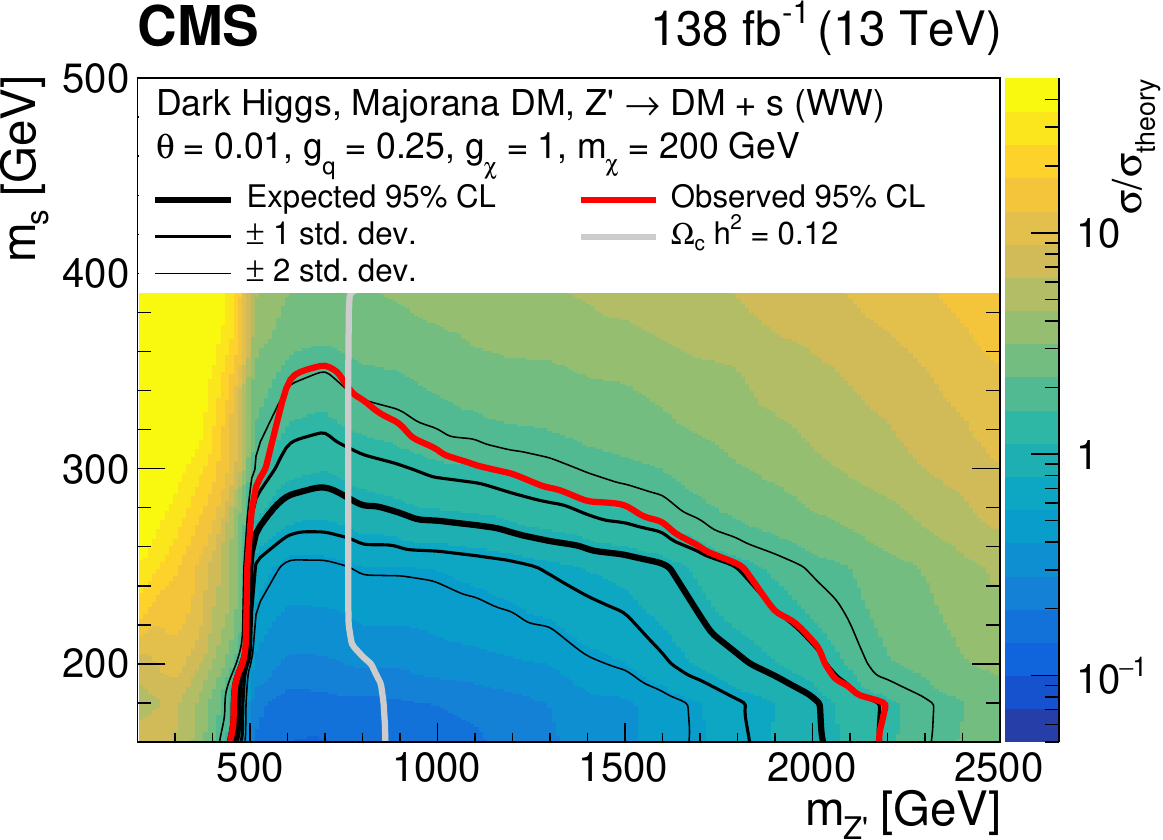}%
\hspace*{0.05\textwidth}%
\includegraphics[width=0.47\textwidth]{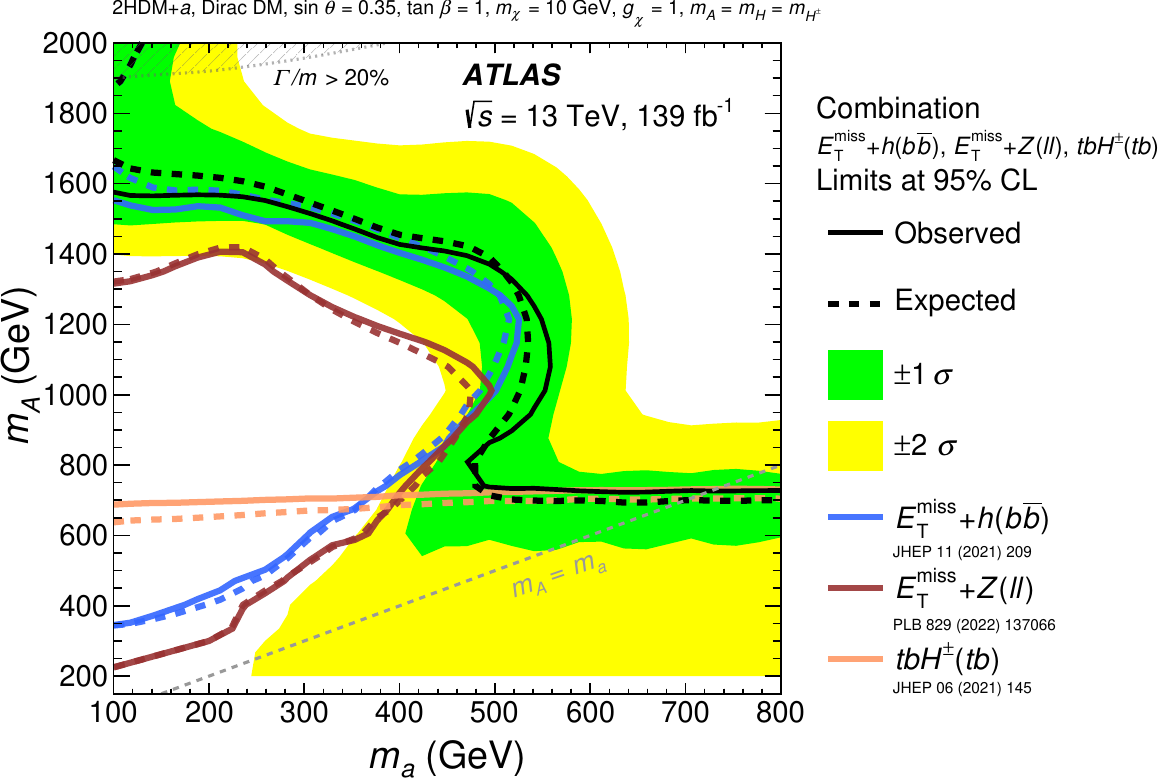}%
\caption{%
    Left: Dark Higgs model limits on the dark Higgs boson mass as a function of the new vector boson mass, obtained by CMS in the search for DM produced in association with a W boson pair~\cite{CMS:EXO-21-012}.
    Right: Limits on the CP-odd Higgs boson mass as a function of the pseudoscalar mediator mass in a 2HDM model extended with DM and a pseudoscalar mediator, obtained by ATLAS in different search channels~\cite{ATLAS:EXOT-2018-64}.
}
\label{fig:3}
\end{figure}

\paragraph{DM produced in association with W boson pair in dark Higgs model (CMS)~\cite{CMS:EXO-21-012}}
Dark sector models comprising a DM candidate particle, a new vector boson, and a dark Higgs boson can provide a relic abundance consistent with cosmological observations even for DM masses where simpler models would be excluded.
This search targets the associated production of the new vector boson together with the dark Higgs boson, where the vector boson decay to DM particles results in large $p_{\mathrm{T}}^{\text{miss}}$ and the dark Higgs boson decay to a W boson pair results in visible SM particles.
Events are selected in the one- and two-lepton channels, where dedicated analysis strategies are applied to separate the signal from the SM background.
Exclusion limits are obtained for DM masses in the range 100--300\,\GeV, vector boson masses in the range 0.2--2.5\,\TeV, and dark Higgs boson masses in the range 160--400\,\GeV.
An example for one fixed value of the DM mass is shown in Fig.~\ref{fig:3}~(left).

\paragraph{DM in two-Higgs-doublet model (2HDM) extended with pseudoscalar mediator (ATLAS)~\cite{ATLAS:EXOT-2018-64}}
A theoretically well-motivated SM extension is the 2HDM that adds an additional complex Higgs doublet and results in one additional scalar, pseudoscalar, and charged Higgs boson each.
Further extending the 2HDM with a pseudoscalar mediator that couples to a DM candidate particle provides a benchmark model with a rich collider phenomenology including signatures not predicted in commonly used simplified DM models.
This interpretation combines three dedicated search channels.
The first two searches target the production of a new heavy Higgs boson, decaying to the invisibly decaying pseudoscalar mediator and either an SM Higgs boson (further decaying to two bottom quarks)~\cite{ATLAS:EXOT-2018-46} or a Z boson (further decaying leptonically)~\cite{ATLAS:HIGG-2018-26}.
The third search targets production of a charged Higgs boson (further decaying to a top and a bottom quark) in association with a top and a bottom quark, thus resulting in a final state with two top and two bottom quarks~\cite{ATLAS:HDBS-2018-51}.
The exclusion limits from the combined interpretation of these three searches are shown in Fig.~\ref{fig:3}~(right).
Additionally, the constraints are compared to those obtained in other searches for new light scalar bosons, highlighting the complementarity between the different search strategies.

\begin{figure}[!ht]
\centering
\includegraphics[width=0.8\textwidth]{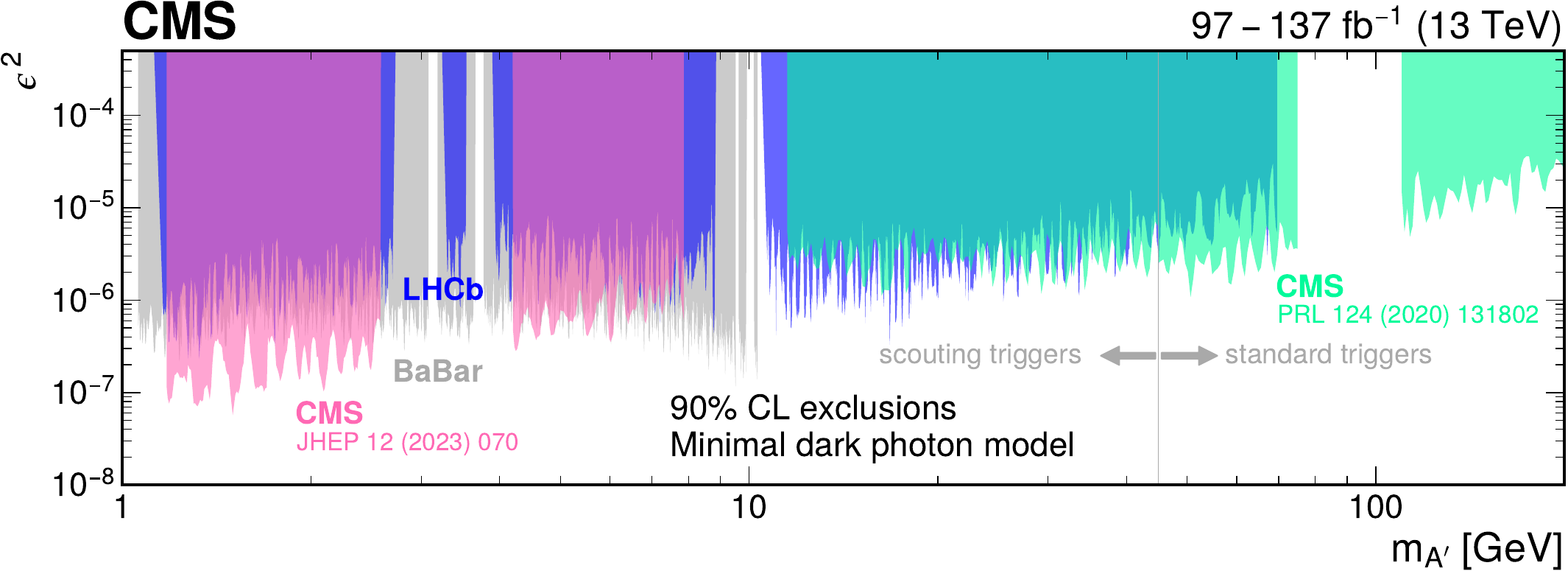}%
\caption{%
    Dark photon limits on the squared kinetic mixing coefficient as a function of the dark photon mass, obtained by CMS with different trigger strategies in different mass ranges~\cite{CMS:Review-DM}.
}
\label{fig:4}
\end{figure}

\paragraph{Low-mass dimuon resonances with scouting triggers (CMS)~\cite{CMS:EXO-21-005}}
A dedicated high-rate trigger stream recording only a small part of the event information reconstructed at the trigger level instead of storing the full raw detector readout as done in standard triggers (``scouting'')~\cite{CMS:Review-Scouting} can be utilized to search for low-mass resonances decaying to muon pairs with much lower transverse momentum than what is typically required for muon selections in new physics searches.
This search selects events with two muons from a common vertex with transverse momenta as low as 4\,\GeV{} to target resonances with masses in the ranges 1.1--2.6 and 4.2--7.9\,\GeV, assuming that the width is much smaller than the experimental resolution.
From sliding mass window fits with analytical signal function and empirical background functions, exclusion limits are evaluated on the dimuon resonance cross section as a function of its mass.
Specific interpretations are performed in a minimal dark photon model that extends the SM with an additional U(1) gauge field that mixes kinetically with the SM hypercharge gauge field, resulting in a massless dark photon.
Exclusion limits for this model are shown in Fig.~\ref{fig:4} and compared to other search results.

\begin{figure}[!ht]
\centering
\includegraphics[width=0.48\textwidth]{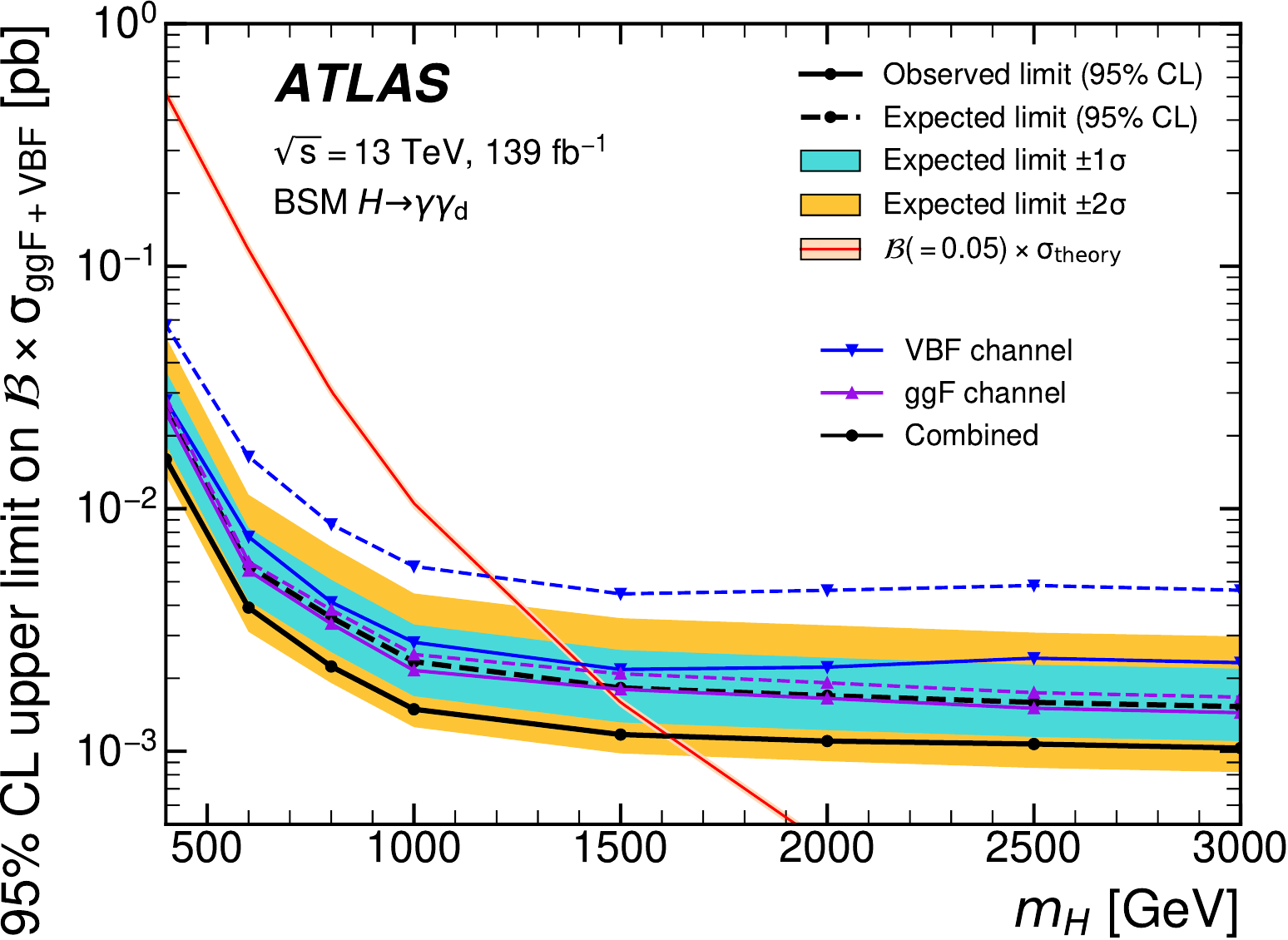}%
\caption{%
    Heavy Higgs boson limits on the production cross section as a function of the heavy Higgs boson mass, obtained by ATLAS in the search for Higgs boson decays to a photon and a dark photon~\cite{ATLAS:EXOT-2023-01}.
}
\label{fig:5}
\end{figure}

\paragraph{Higgs boson decays to photon and massless dark photon (ATLAS)~\cite{ATLAS:EXOT-2023-01}}
A dark photon can also couple to the Higgs sector, resulting in signatures where the SM Higgs boson or a new heavy Higgs boson from an extended Higgs sector can decay to an SM photon and an undetectable dark photon.
This interpretation considers three different Higgs boson production modes (gluon-gluon fusion without additional SM particles, Z-boson associated production, and vector boson fusion production with two additional forward jets), where the subsequent decay to an SM photon and a dark photon results in three distinctive signatures: $\gamma$\,+\,$p_{\mathrm{T}}^{\text{miss}}$~\cite{ATLAS:EXOT-2018-63}, $\mathrm{Z}(\ell\ell)$\,+\,$\gamma$\,+\,$p_{\mathrm{T}}^{\text{miss}}$~\cite{ATLAS:HDBS-2019-13}, and two forward jets\,+\,$\gamma$\,+\,$p_{\mathrm{T}}^{\text{miss}}$~\cite{ATLAS:EXOT-2021-17}.
Dedicated searches for these signatures are combined and interpreted in the two scenarios.
In the SM Higgs boson scenario, branching fractions larger than 1.3\% are excluded.
In the heavy Higgs boson scenario, masses between 400\,\GeV{} and 3\,\TeV{} are considered, and the resulting exclusion limits are shown in Fig.~\ref{fig:5}.

\section{Summary}

The large amount of recorded data and the high collision energy of the 2015--2018 data-taking period at the LHC is used by the ATLAS and CMS Collaborations to search for new physics beyond the SM with various signatures.
Several examples of searches for promptly produced weakly interacting particles have been presented, targeting dark meson, heavy neutral lepton, dark matter, and dark photon models.
By developing dedicated analysis strategies, often employing machine-learning techniques, the results significantly extend the reach of exclusion limits or provide sensitivity to new models for the first time.
A new LHC data-taking period has started in 2022 at a slightly higher energy of 13.6\,\TeV, and by now the delivered integrated luminosity has exceeded that of the 2015--2018 data set.
Further advancements in new-physics searches can be expected, both from further refined analysis techniques and from the new data set.

\acknowledgments

The author acknowledges support from the Research Foundation Flanders (FWO) as a senior postdoctoral fellow fundamental research (grant number 1287324N).

\newcommand{\arxiv}[2]{\href{http://arxiv.org/abs/#1}{arXiv:#1} (#2)}
\newcommand{\cds}[1]{\href{http://cds.cern.ch/record/#1}{CDS:#1}}
\newcommand{\doi}[2]{\href{http://doi.org/#1}{#2}}
\newcommand{\bibtitle}[1]{\emph{#1},}

\end{document}